\newcommand{\bb}{\begin{equation}}
\newcommand{\en}{\end{equation}}
\documentclass[epj]{svjour}
\usepackage{dcolumn}
\usepackage{amsmath2000}
\usepackage{latexsym}
\usepackage{graphics}
\begin{document}
\title{Phase Transitions in Lyotropic Nematic Gels}
\author{D. Lacoste \thanks{\emph{Permanent address:} Laboratoire de
Physico-Chimie Th\'eorique, ESPCI, 10 rue Vauquelin, 75231 PARIS
Cedex 05; e-mail: david@turner.pct.espci.fr}, A.W.C. Lau  and T.C.
Lubensky}
\institute{Department of Physics and Astronomy, University of
Pennsylvania, Philadelphia, PA 19104}
\date{Received: date / Revised version: date}
%
\abstract{In this paper, we discuss the equilibrium phases and
collapse transitions of a lyotropic nematic gel immersed in an
isotropic solvent. A nematic gel consists of a cross-linked
polymer network with rod-like molecules embedded in it. Upon
decreasing the quality of the solvent, we find that a lyotropic
nematic gel undergoes a discontinuous volume change accompanied by
an isotropic-nematic transition.  We also present phase diagrams
that these systems may exhibit. In particular, we show that
coexistence of two isotropic phases, of two nematic phases, or of
an isotropic and a nematic phase can occur.
} 
\maketitle

\section{Introduction}
\label{sec:introduction}

A polymer gel consists of flexible polymer chains that are
cross-linked to form a three-dimensional network \cite{Flory}
whose volume in response to solvent changes can swell by factors
as great as 1000 or more and undergo discontinuous changes along a
coexistence line terminating in a liquid-gas like critical point
\cite{tanaka1}. Dispersions of long rigid rods undergo a
transition from an isotropic to a lyotropic nematic phase with
increasing rod volume fraction \cite{onsager,fraden}. In this
paper, we will explore the possibility of creating lyotropic
nematic phases in polymer gels with rigid rods either dispersed in
their open spaces or forming parts of their constituent polymer
chains. Our basic premise is that decreases in solvent quality
contract the gel, thereby increasing rod volume fraction and
inducing a transition to a nematic phase.  A lyotropic nematic gel
has the same spontaneous broken rotational symmetry and
macroscopic elastic properties as a thermotropic nematic or
elastomer \cite{FinKoc81,deGennes_gels,terentjev2,lubensky}.  As a
result, it will exhibit the characteristic ``soft" elasticity
\cite{lubensky,golubovic,olmsted,WarBla94} of a nematic gel
whereby the shear modulus for shears in planes containing the
anisotropy axis vanishes and stresses up to critical distortions
vanish for certain elongational and compressional strains.  This
``soft" elasticity along with unusually large coefficients of
thermal expansion suggest interesting technological applications
for thermotropic nematic or other liquid crystalline gels
including artificial muscles \cite{hebert,ratna}, actuators
\cite{zentel}, and electro-optical devices \cite{electro-opt}.
Lyotropic nematic gels might find use in similar applications.  It
is also imaginable that nature has found ways to use biological
versions of these materials to control elastic properties of
cells.

A theory for lyotropic nematic gels should be sophisticated enough
to treat both the volume-collapse transition \cite{doi,escobelo} of
an isotropic gel and the development of nematic order.
Tanaka \cite{tanaka1} explained the former phenomena using Flory's
theory for gels \cite{Flory,deGennes,flory3}, which combines the
Flory-Huggins theory of polymer solutions with the Flory theory
for the entropic elasticity of isotropic elastomers. Brochard
\cite{brochard} considered gels in thermotropic liquid crystal solvents
and concluded that a gel crosslinked in the isotropic phase of the
solvent collapses when the solvent undergoes a transition from the
isotropic to the nematic phase.  Warner and
Wang \cite{warner1} modified Flory's theory to treat anisotropic
gels using the neoclassical theory \cite{WarBla94} for the
elasticity of anisotropic rather than isotropic elastomers. They
then used the phenomenological Maier-Saupe theory \cite{deGPro93}
to describe the development of nematic order. Coupling between
elasticity and nematic order is provided by the dependence of
anisotropic step-lengths in the neoclassical elasticity theory on
the nematic order parameter. They found a rich phase diagram, with
temperature as a control parameter, in which coexistence of
nematic and isotropic phases and a discontinuous uptake of solvent
are possible. In this paper, we follow Warner and Wang's treatment
of gel elasticity, but we use the Onsager theory \cite{onsager},
which provides a quantitatively accurate description for the
isotropic-to-nematic (I-N) transition in dispersions of hard
rods \cite{fraden}, rather than Maier-Saupe theory to describe the
development of nematic order. Our theory provides a more accurate
treatment of the strongly first-order I-N transition of
dispersions of hard-rods than does one based on the
phenomenological Maier-Saupe theory, which is best suited for a
description of the vicinity of the transition in thermotropic
systems.

\begin{figure}
{\par\centering
\resizebox*{3in}{2.2in}{\rotatebox{0}{\includegraphics{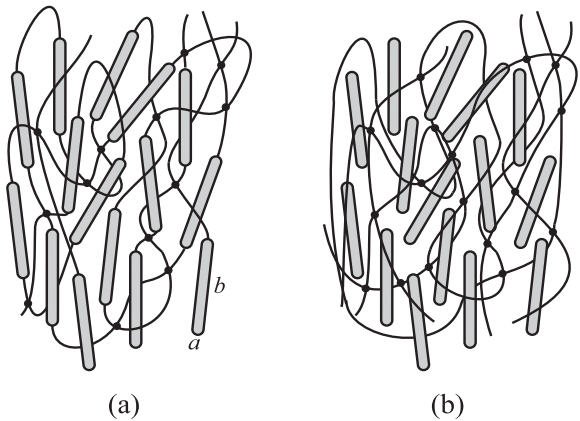}}}
\par}
\caption{Representation of the nematic gels referred to as (a)
model 1 and (b) model 2. In both cases, rods of diameter $a$ and
length $b$ are embedded in a crosslinked polymer network. In model
1, the rods are part of the network which consists of crosslinked
polymer chains composed of rigid rods and flexible spacers. The
thickness of the spacers is equal to the diameter of the rods, and
the rods are much smaller than the average distance between
crosslinks. In model 2, the independent rigid rods are dispersed
in the free spaces of a crosslinked flexible polymer network.
There is no particular constraint on the distance between
crosslinks and on the size of the chains which constitute the
network.} \label{Fig:picture}
\end{figure}

We consider two models depicted schematically in Fig.
\ref{Fig:picture}: model 1 in which the polymer chains
constituting the gel consist of rigid rods connected by flexible
segments and model 2 in which rigid rods are dispersed in the
fluid volume between polymer segments. Though we introduce certain
not completely controlled approximations to be discussed below, we
believe that these models provide a realistic semi-quantitative
description of lyotropic nematic gels. In both models, we find as
expected a first order I-N transition accompanied by a
discontinuous change in volume.  As in the systems considered
by Brochard \cite{brochard}, the development of nematic order is
accompanied by a decrease in the gel volume.  In model 2, we also
find lines along which either two isotropic or two nematic phases of
differing densities coexist. These lines terminate in mean-field
liquid-gas-like critical points, which are expected to retain
their mean-field character even when fluctuations beyond
mean-field theory are included because of effective long-range
forces induced by a non-vanishing shear modulus \cite{golubovic}.

This paper is organized as follows: In the next section (Sec.
\ref{sec:model}), we develop the model free energy for both models
1 and 2 including the orientational free energy of rods described
by the Onsager theory. In Sec. \ref{sec:equil_swelling}, we
construct the phase diagrams for models 1 and 2.  We conclude with
a brief summary.

\section{Free energy of nematic gels}
\label{sec:model}

In this section, we will derive the free energy for nematic gels.
This free energy includes entropic and enthalpic contributions
that are present in uncrosslinked polymer solutions, elastic
contributions arising from crosslinking, and contributions arising
from orienting rigid rods, and it depends  the volume fractions of
polymer, solvent, and rods. We denote by $\phi$ the total volume
fraction of the network, which contains flexible polymers of
volume fraction $\phi_P$ and rods of volume fraction $\phi_R$, so
that
\begin{equation}\label{phi_network}
\phi=\phi_P+\phi_R.
\end{equation}
The solvent volume fraction is thus $1-\phi$. Under the assumption
that the rods are confined within the network as it deforms, the
ratio of the concentration of the rods to that of the network must
be a constant, which we denote by
\begin{equation}\label{beta}
\beta=\frac{\phi_R}{\phi}.
\end{equation}
In model 2, we have not considered the possibility of an
equilibrium of the mobile rods between the inside and the outside
of the gel as done in Ref.~\cite{brochard} for flexible rods
chemically identical to the polymer chains of the network. Instead
by using Eq.~\ref{beta} with a fixed $\beta$, we have assumed that
a constant number of rods stay trapped within the gel as the gel
changes its volume. It is possible that the rods stay confined in
the polymer network either for kinetic reasons or because of
entanglements within the network.

To describe nematic gels immersed in a solvent, we assume that the
total free energy can be written as a sum of three contributions
\begin{equation}\label{def_free_energy}
{\cal F}={\cal F}_{mix}+{\cal F}_{el}+{\cal F}_{rod},
\end{equation}
where ${\cal F}_{mix}$ is the free energy of mixing of the gel
with its solvent, ${\cal F}_{el}$ is the elastic free energy of
the gel network, and ${\cal F}_{rod}$ is the free energy of the
rods. Within the Flory-Huggins theory, ${\cal F}_{mix}$ may be
written as \cite{Flory,tanaka1,warner1} \bb \label{F_mix} {\cal
F}_{mix} = { V k_B T \over v_c } f_{mix}, \en where $V$ is the
volume of the gel and $v_c$ is the volume of a lattice site, which
is assumed to be $v_c=a^3$, where $a$ is the monomer diameter.
$f_{mix}$ is the free energy per lattice site given by
\begin{eqnarray}
\label{f_mix} f_{mix} &=&( 1-\phi) \ln ( 1- \phi) + \chi\,\phi_P
(1-\phi) + \chi_R\, \phi_R (1-\phi) \nonumber \\
&\phantom{+}& \, + \,\chi_P\,\phi_R \phi_P,
\end{eqnarray}
where $\chi$, $\chi_R$, and $\chi_P$ are the Flory-Huggins
parameters that characterize, respectively, the polymer-solvent,
the rod-solvent, and the rod-polymer interactions at the level of
the second virial approximation. The first term in
Eq.~(\ref{f_mix}) represents the entropy of the solvent. Note that
the translational entropies of the polymers, $\phi_P \ln \phi_P $,
and rods, $\phi_R \ln \phi_R$, are absent in Eq.~(\ref{f_mix})
because the chains, which include the rods, are attached to the
network \cite{deGennes}. In order for the Onsager theory to be
valid, the solvent quality must be good for the rods so that their
mutual interactions are well described by the excluded volume
interaction.  For simplicity, we assume that $\chi_R = \chi_P =0$
corresponding to an athermal solvent for the rods and absence of a
direct rod-polymer interactions. Thus, the single parameter
$\chi$, which measures the strength of interaction between polymer
and solvent, controls the solvent quality. It is expected to vary
with temperature as $\Theta/2T$, where $\Theta$ is the
theta-temperature separating good-solvent ($T>\Theta$) from
poor-solvent ($T<\Theta$) behavior.

The elastic free energy ${\cal F}_{el}$ introduced by Warner {\it
et al.} \cite{WarBla94} describes the energy change associated
with a deformation of a nematic elastomer characterized by
homogenous strains parallel $\lambda_\parallel$ and perpendicular
$\lambda_\perp$ to the director (along which the rods are
aligned).  It may be written as \bb
\label{F_el}
{\cal F}_{el} =
{1 \over 2} k_B T N_c \left [ \lambda_\parallel^{2} \, {\ell_0
\over \ell_\parallel} + 2\,\lambda_\perp^{2}\,{\ell_0 \over
\ell_\perp} + \ln \left ( \frac{\ell_\parallel
\ell_\perp^2}{\ell_0^3} \right ) \right ],
\en
where $N_c$ is the
total number of network strands, $\ell_0$ is the step length of
the polymer in the isotropic phase in which we assume the system
was cross-linked, and $\ell_\parallel$ and $\ell_\perp$ are,
respectively, the step lengths parallel to and perpendicular to
the director in the nematic phase. Strains $\lambda$ arise
spontaneously due to the anisotropy of the network, and they are
measured with respect to an isotropic state characterized by
$\ell_0$. Note that $\ell_\parallel$ and $\ell_\perp$ are in
general functions of the de Gennes-Maier-Saupe \cite{deGPro93}
order parameter $Q_{ij}$, which describes the orientational
ordering of the rods, and they reduce to $\ell_0$ if $Q_{ij} = 0$.
Equation (\ref{F_el}) is based on the following assumptions: (i)
all chains are sufficiently long that their end-to-end vectors may
be described by an anisotropic Gaussian distribution, and (ii)
this distribution transforms affinely under macroscopic
deformation. Equation (\ref{F_el}) has been used to account for
many properties of nematic elastomers
\cite{terentjev2,WarBla94,terentjev}.

Equation (\ref{F_el}) differs from the Flory expression for the
elastic free energy of a gel \cite{Flory,flory3}, which contains
an additional strain-dependent term of the form $\ln \left(
\lambda_\parallel^{2} \lambda_\perp \right)$. While Tanaka {\em et
al.} \cite{tanaka1} and Matsuyama {\em et al. } \cite{matsuyama}
have used the Flory theory with this term included to explain the
collapse of isotropic gels, we believe (and later check) that this
term does not play a significant role in nematic gels. Therefore,
we do not include this term in the discussion that
follows \cite{warner1,deGennes,edwards}. It is, nevertheless,
interesting to note that this term is expected to be absent in the
so-called phantom model of rubber elasticity \cite{edwards}. This
model essentially ignores the effect of entanglements of the
chains and thermal fluctuations of the junctions points and is
believed to be valid in the dilute regime. In more refined models,
however, where these effects are properly taking into account,
there may be additional strain dependent terms contributing to
Eq.~(\ref{F_el}) \cite{everaers}.

Unlike thermotropic nematic elastomers, which are essentially
incompressible, lyotropic nematic gels can undergo large volume
changes because of their ability to take up solvent \cite{treolar}.
We measure volume changes by the swelling ratio
$V/V_0=\phi_0/\phi$, where $V_0$ is the network volume (and
$\phi_0$ its volume fraction) in the reference state, and $V$ is
the network volume (and $\phi$ its volume fraction) in any
particular state of gel. The strains are related to the swelling
ratio by
\begin{equation}
\label{change_vol}
\lambda_\parallel
\lambda_\perp^2=\frac{V}{V_0}=\frac{\phi_0}{\phi}.
\end{equation}
This relation allows us to express the elastic free energy
Eq.~(\ref{F_el}) in terms of $\lambda_\parallel$ only. Minimizing
Eq.~(\ref{F_el}) with respect to $\lambda_\parallel$, we obtain
\begin{equation} \label{lambda_equil}
\lambda_\parallel=\left( \frac{\phi_0 \ell_\parallel}{\phi
\ell_\perp} \right)^{1/3}
\end{equation}
and
\begin{eqnarray}
f_{el}=\frac{{\cal F}_{el} v_c}{k_B T V} &=& {3 \over 2}\, \nu_c \,
\left( { \phi \over  \phi_0 }\right)^{1/3} \left [ { \ell_0 \over
\left( \ell_\parallel \ell_\perp^2 \right)^{1/3}}
\vphantom{ - \left( { \phi \over  \phi_0 }\right)^{2/3} \,
\ln { \ell_0 \over \left( \ell_\parallel \ell_\perp^2 \right)^{1/3}}}
\right. \nonumber \\
&\phantom{-}& \,\,\,\,\left. - \left( { \phi \over  \phi_0 }\right)^{2/3} \,
\ln { \ell_0 \over \left( \ell_\parallel \ell_\perp^2 \right)^{1/3}}
\right],
\label{F_el_min}
\end{eqnarray}
where $\nu_c=N_c v_c/V_0$. Note that Eq. (\ref{F_el_min}) is
slightly more general than the one used by Warner et al. \cite{warner1},
where the particular case of $\phi_0=1$, {\em
i.e.} a dry gel, is considered. As pointed out in Ref. \cite{tanaka1},
in order for an isotropic gel to undergo collapse
it is crucial that $\phi_0 \ll 1$, and we expect a similar
conclusion to apply to nematic gels.

The free energy ${\cal F}_{rod}$ describes the physics of the rods
embedded in the gel. Unlike previous treatments of nematic
elastomers and gels, which focus mostly on thermotropic systems,
our current treatment focusses on lyotropic gels. We employ
Onsager theory \cite{onsager} to describe the isotropic-nematic
transition that occurs upon increasing rod volume fraction via
osmotic compression of the gel. The free energy obtained by
Onsager is \cite{onsager}\begin{eqnarray}
{{\cal F}_{rod} \over k_B T} &=& N_R \int d^2{\bf u}\,
\Psi({\bf u})\ln \left[ \frac{4\pi
\Psi({\bf u}) N_R}{Ve} \right] \nonumber \\
&& + \frac{a b^2 N_R^2}{V} \int
d^2{\bf u}\,d^2{\bf u}'\,\Psi({\bf u})\Psi({\bf u}') |{\bf u} \times {\bf u}'|,
\label{F_rod}
\end{eqnarray}
where ${\bf u}$ is the unit vector
specifying rod direction, $N_R$ is the number of rods in the
network with length $b$ and diameter $a$ and $\Psi({\bf u})$ is
the orientational distribution function. The first term in Eq.
(\ref{F_rod}) is the entropy of the rods, and the second takes
into account the excluded-volume interaction between the rods at
the level of the second virial approximation, which becomes exact
in the limit of infinitely long rods.  Equation (\ref{F_rod}) is the
Onsager free energy for rods in an isotropic solvent, and not for
rods either attached to or dispersed in a network.  It clearly
overestimates both rotational and translational entropy of rods since the
network will introduce steric constraints whose effects are difficult to
estimate.  Using the Onsager free energy for the rods rather than one that
takes full account of constraints imposed by the network is the principal
approximation of our theory.  As discussed below, we will treat the rod
translational entropy differently in models 1 and 2.

Even in the absence of the coupling between the strains and
nematic order, minimizing the free energy ${\cal F}_{rod}$ yields
an intractable integral equation for $\Psi({\bf u})$.  We,
therefore, follow Onsager \cite{onsager,warren} and introduce a
trial distribution function of the form \bb
\Psi({\bf u}) = {\alpha \cosh( \alpha {\bf u} \cdot {\bf c} )
\over 4 \pi \sinh \alpha},
\en
where $\alpha$ is a variational parameter and ${\bf
c}$ is an arbitrary unit vector. Thus, the two integrals in Eq.
(\ref{F_rod}) can be performed analytically:
\begin{eqnarray}
\label{sigma}
\sigma(\alpha)&=&\int d^2{\bf u}\, \Psi({\bf u})\ln \left[ 4\pi
\Psi({\bf u}) \right] \nonumber \\
&=& \ln \left( \alpha \coth \alpha \right) -1 +
\frac{\arctan \left( \sinh \alpha \right)}{\sinh \alpha} \\
\label{rho}
\rho(\alpha)&=&\frac{4}{\pi} \int d^2{\bf u}\, d^2{\bf
u}'\,\Psi({\bf u})\Psi({\bf u}') |{\bf u} \times {\bf
u}'| \nonumber \\
&=& \frac{2I_2(2\alpha)}{\sinh^2 \alpha},
\end{eqnarray}
where $I_2(x)$ is a modified Bessel function.

Using these results and Eq. (\ref{beta}), we can calculate the rod
free energy per site.  In model 1, the rods are locked to the
network, and there is no independent translational entropy
associated with them, although there is rod translational entropy
in $f_{el}$ arising from fluctuations of the polymer-rod segments
constituting the network.  In model 2, rods are not attached to
the network, and we assume for simplicity that they are free to
diffuse throughout the volume of the gel not occupied by polymer
segments and that they therefore have a translational entropy per
site of $\phi_R \ln \phi_R$.  In real systems, rods may in be
sterically confined to finite regions of the gel in which case,
our model overestimates the rod translational entropy. The free
energies $f_{rod}^1$ and $f_{rod}^2$ are thus, respectively,
\begin{eqnarray}
f_{rod}^1& =& \frac{{\cal F}_{rod} v_c}{k_B T V} = \frac{4a}{\pi
b} \phi \beta  \sigma(\alpha)  + \frac{4}{\pi} \phi^2 \beta^2
\rho(\alpha)
\label{F_rod_2}\\
f_{rod}^2 & =& f_{rod}^1 + {4 a \over \pi b} \phi \beta \ln \phi
\label{F_rod_1}
\end{eqnarray}
up to an irrelevant constant in the chemical potential. Note that
the term proportional to $\phi \ln \phi$ represents the
translational entropy of the rods. The order parameter of the I-N
transition is defined by the tensor \cite{tom}
\begin{equation}\label{tensorQ}
Q_{ij}=\int d^2{\bf u}\, \Psi({\bf u}) \left[ u_i u_j -
\frac{1}{3} \delta_{ij} \right].
\end{equation}
This tensor can be diagonalized to yield \[
Q={\rm Diag} \left[ -S/3,-S/3,2S/3 \right],
\]
where $S$ is the scalar order parameter given by
\begin{eqnarray}
S(\alpha) &=& \frac{1}{2} \int d^2{\bf u}\, \Psi({\bf u}) \left[ 3
({\bf u} \cdot {\bf c})^2 -1 \right] \nonumber \\
&=& 1-\frac{3}{\alpha} \left(
\coth \alpha - \frac{1}{\alpha} \right).
\label{S_alpha}
\end{eqnarray}
Note that $S(\alpha)$ vanishes in the isotropic phase when
$\alpha=0$ and that $S(\alpha) > 0$ in the nematic phase when
$\alpha>0$.

To complete our description of nematic gels, we must specify the
relation between the anisotropy of the polymer chains and the
nematic order. We define a step length tensor by $\ell_{ij} =
\langle R_i R_j \rangle / N \ell_0$, which conveniently represents
the anisotropy of the chains, where ${\bf R}$ is the end-to-end
vector of the chains between two crosslinks \cite{muthukumar} and
$N$ is the total number of elements in these chains. The
eigenvalues of $\ell_{ij}$ are $\ell_\parallel$ and $\ell_\perp$,
which enter into the elastic free energy Eq. (\ref{F_el}). In
general, $\ell_{ij}$ is some unknown function of the nematic order
parameter $Q_{ij}$, which may be model dependent. The simplest
form of this function allowed by symmetry is \cite{terentjev2,lubensky}
\begin{equation}
\label{freely-jointed}
\frac{\ell_{ij}}{\ell_0}=\delta_{ij}+\frac{1}{2} \eta \, Q_{ij},
\end{equation}
with an unknown coupling parameter $\eta$, which depends on the
network structure and which will be different for models 1 and 2.
Such a linear relationship is well verified experimentally for
nematic elastomers \cite{finkelman}.

In the case of model 1, we assume that there are $\gamma N$ rods
of length $b$ and volume $\pi a^2 b/4$, and $(1-\gamma)N$ spacers
of size $a$ and volume $a^3$, which are distributed randomly in a
polymer chain, forming a main-chain liquid crystalline polymer \cite{muthukumar}.
Using the freely-jointed chain (FJC) model \cite{warner1,matsuyama}, we find that
\begin{equation}
\label{eta}
\eta = \eta_1=\frac{6 \gamma (b/a)^2}{1-\gamma + \gamma (b/a)^2},
\end{equation}
where $\gamma$ is related to $\beta$ for model 1 in the following
way
\begin{equation}\label{def_gamma}
\beta=\frac{\pi b \gamma}{4a \left(1-\gamma \right)+\pi b \gamma}.
\end{equation}
Note that Eq.~(\ref{eta}) is exact for the FJC model and is valid
for any form of the distribution function $\Psi({\bf u})$.

For model 2, on the other hand, the FJC result is not expected to
apply since the rods are not physically connected to the network
chains. However, it is reasonable that for a sufficiently low
concentration of the rods, Eq.~(\ref{freely-jointed}) should still
hold with an unknown coefficient $\eta$, which we take arbitrarily
to be $\eta = \eta_2 = 6$ (the maximum value of $\eta$ in model
1). Using Eqs.~(\ref{S_alpha}) and (\ref{freely-jointed}), we have
\begin{eqnarray} \label{parallel}
{\ell_\parallel(\alpha)\over \ell_0} &=& 1+ {\eta  \over 3} S(\alpha), \\
{\ell_\perp(\alpha)\over \ell_0} &=& 1- {\eta  \over 6} S(\alpha).
\label{perp}
\end{eqnarray}
In both models $S>0$ in equilibrium in the nematic phase,
$\ell_\parallel
> \ell_\perp$, and the gel adopts a prolate shape.  Generalization
to allow an oblate phase are not difficult but will be left for
another study.  Our final free energy per site for model 1 is
$f_{mix} + f_{el} + f_{rod}^1$ with the parameter $\eta$ given by
Eq.~(\ref{eta}) while that for model 2 is $f_{mix} + f_{el} +
f_{rod}^2$ with $\eta = \eta_2 = 6$.  We analyze the phase
diagrams arising from these free energies in the next section.

\section{Equilibrium swelling of a nematic gel}
\label{sec:equil_swelling}

In this section, we discuss the equilibrium properties of a
nematic gel. The orientational order of the nematic gel is
controlled by the parameter $\alpha$, which is determined by the
condition that the total free energy per site $f =
f_{mix}+f_{el}+f_{rod}$ is a minimum at $\alpha = \alpha^*(\phi)$.
The constraint $\phi \leq 1$ implies that $\alpha^*(\phi) \leq
\alpha_{max}$, where $\alpha_{max}$ is the value of $\alpha^*$
obtained for $\phi=1$. At $\alpha=\alpha_{max}$ the maximum
possible value of scalar order parameter $S$ is reached. Note that
the orientational order is governed only by $f_{rod}$ and $f_{el}$
and not by $f_{mix}$. The elastic free energy couples to the
nematic order through $\ell_\parallel(\alpha)$ and
$\ell_\perp(\alpha)$ as given in Eqs.~(\ref{parallel}) and
(\ref{perp}). As pointed out in Ref.~ \cite{warner1}, although the
contribution from $f_{el}$ is not negligible, the presence of the
cross-linked network does not qualitatively alter the nematic
properties of the rods. In particular, if the volume fraction
$\phi$ is small, $f$ has only one minimum at $\alpha =0$, which
corresponds to the isotropic state.  If $\phi > \phi_1$, another
minimum appears at positive $\alpha$, which corresponds to the
nematic state. If $\phi > \phi_2$, the minimum at $\alpha =0$
disappears and there is only one minimum at a positive $\alpha$.
Therefore, the isotropic state is only possible for $ 0 \leq \phi
< \phi_2$ and the nematic state for $\phi_1 < \phi \leq 1$.  These
features are also found in Onsager theory for lyotropic liquid
crystals. However, in nematic gels, the equilibrium phases are
more interesting due to the coupling of elastic strains and
nematic order.

The chemical potential $\mu$ and the osmotic pressure $\Pi$ are
two quantities characterizing a nematic gel. They are
respectively, given by
\begin{eqnarray}
\mu &=& \left. \partial f /\partial \phi \right|_{\alpha=\alpha^*(\phi)}, \\
\label{osmotic_pressure}
\Pi &=& -\frac{\partial {\cal F}}{\partial
V}=-\frac{k_B T}{v_c} \left[ f - \phi \frac{\partial f}{\partial
\phi} \right]_{\alpha=\alpha^*(\phi)}.
\end{eqnarray}
In the nematic phase, their dependance on $\phi$ includes
contribution from the dependance of $\alpha^{*}(\phi)$ on $\phi$.
The chemical potential governs the uptake of the solvent molecules
and the osmotic pressure the mechanical stability of the gel.  If
$\Pi \geq 0 $, the gel is stable with the surrounding solvent.  If
$\Pi < 0$ the gel becomes unstable in the absence of an external
applied stress, and consequently it shrinks and separates itself
from its fluid until it reaches $\Pi = 0$. Thus, the equilibrium
swelling of the gel follows the volume curve $\phi^*(\chi)$
defined by the condition $\Pi = 0$, which is equivalent to
minimizing the total free energy ${\cal F}$ with respect to
$\phi$. Note that in analogy with the phase diagram of the
liquid-gas model, it is interesting to consider isobars
$\phi^*(\chi; \Pi_0)$ describing a gel to which a constant
pressure $\Pi_0$ is applied \cite{Flory}. In the following, we
focus mainly on the situation with no applied stress $\Pi_0=0$,
and we will not consider situations in which $\Pi_0<0$, which can
only be reached by attaching the outer surfaces of the gel to
movable walls that can be subjected to tension. In order to reach
an equilibrium state of the gel, the external condition, such as
the temperature, must change very slowly. On the other hand, if
the temperature, and thus $\chi$, changes rather rapidly, the gel
does not have time to relax, and its concentration and volume are
thus fixed. In this case, macroscopic phase separation \cite{doi}
may take place if $\phi_A < \phi < \phi_B $, which means that an
isotropic (or nematic) phase of volume fraction $\phi_A$ coexists
with an isotropic (or nematic) phase of volume fraction $\phi_B$
within the same gel, where $\phi_A$ and $\phi_B$ are determined by
the common tangent construction \cite{safran}: (i) the osmotic
pressure in the two phases are equal: $\Pi(\phi_A) = \Pi(\phi_B)$
and (ii) the chemical potential are equal: $\mu(\phi_A) =
\mu(\phi_B)$. We emphasize that in Onsager theory, the I-N
transition is independent of temperature and depends only on the
volume fraction of the rods.  Therefore, in our model for nematic
gels, $\chi$ is the only parameter that depends on the temperature
($\chi \sim 1/T$). This is in contrast to previous
studies \cite{warner1} where the temperature dependence comes from
the Maier-Saupe interaction parameter and $\chi$ is consequently
assumed to be a constant independent of temperature. We present in
what follows the phase diagrams for models 1 and 2 in the
coordinates $\phi$ vs. $\chi$ or $\Pi$ vs. $\chi$ for fixed values
of the parameters $\nu_c$, $\beta$, $b/a$, and $\phi_0$.

\subsection{Nematic gels with rods linked to the network: model 1}

\label{model2}
\begin{figure}
{\par\centering
\resizebox*{2.5in}{2.5in}{\rotatebox{0}{\includegraphics{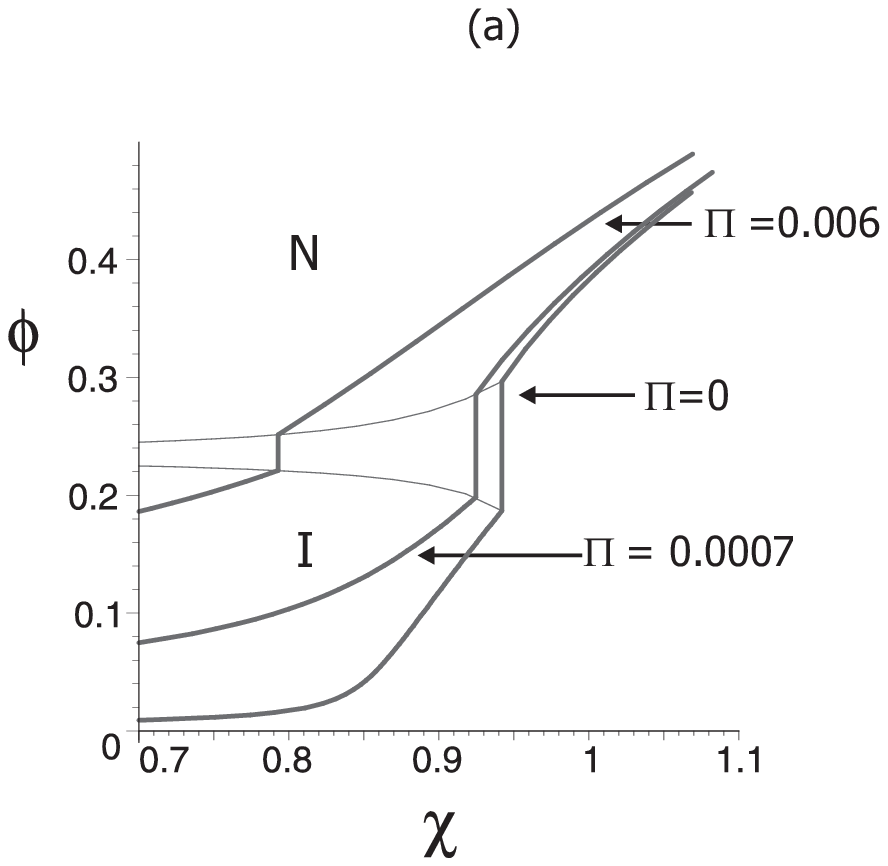}}}
\resizebox*{2.5in}{2.5in}{\rotatebox{0}{\includegraphics{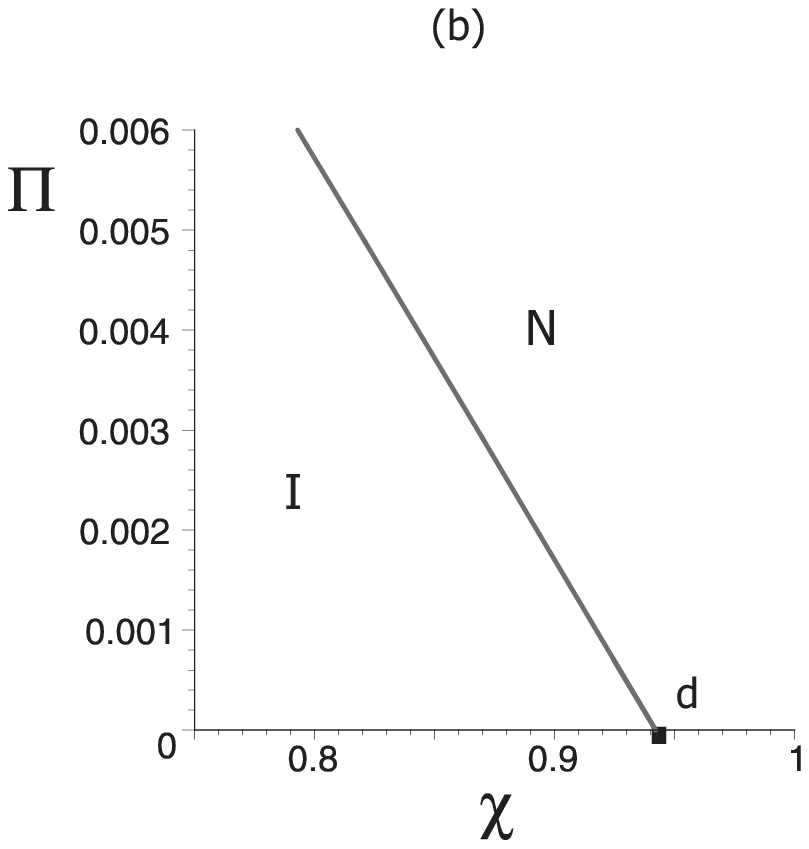}}}
\par}
\caption{Phase diagram for model 1 in the coordinates: (a) volume
fraction of the network $\phi$ vs. Flory-Huggins parameter $\chi$
and (b) osmotic pressure $\Pi$ vs. $\chi$. Thick solid lines are
isobars of constant osmotic pressure for $\Pi=0.006$, $\Pi=0$ and
$\Pi=0.0007$. Thin solid lines delimit the isotropic-nematic
coexistence: the lower line $\phi_A(\chi)$ is the boundary of the
isotropic phase (I) and the upper line $\phi_B(\chi)$ is the
boundary of the nematic phase (N). The isobar $\Pi=0$ is called
the volume curve, and delimits the region of absolute stability of
the gel in the absence of applied stress. The volume curve is
discontinuous (collapse) at $\chi^* \simeq 0.94$, which
corresponds to the point $d$ of coordinate ($\chi^*$,$0$) in (b).
The parameters are $b/a=50$, $\beta=0.284$, $\nu_c = 7 \cdot
10^{-6}$ and $\phi_0 = 5 \cdot 10^{-3}$.} \label{Fig1}
\end{figure}

In model 1, the rods are part of the network in which a polymer
chain consists of rigid rods and flexible spacers. Note that the
parameters $\phi_0, \nu_c, \beta, b/a$ and $\ell_0$ are not all
independent. This is because we have assumed that the deformation
of the network is affine in deriving the elastic free energy
Eq.~(\ref{def_free_energy}). This in turn implies that there is
only one characteristic length scale of the network, the
end-to-end separation of its constituent chains, and this length
must be of the order of the average distance between cross-links
\cite{geissler}. This condition imposes that $\nu_c \sim
({a/\ell_0})^3 N^{-3/2}$ and $\phi_0 \sim ( a/\ell_0)^2 N^{-1/2}$,
where $N$ is the total number of elements (spacers+rods) in the
chains between cross-links \cite{tanaka1}.  Note that only when
$N\gg1$, the average distance between cross-links is larger than
the rod length, which is required to justify a posteriori
Eq.~(\ref{beta}).  For example, for $b/a=50$, $\beta=0.284$ and
$N=100$, we find that $\nu_c \simeq 7 \cdot 10^{-6}$ and $\phi_0
\simeq 5 \cdot 10^{-3}$. These values are used below to construct
the phase diagram.

\begin{figure}
{\par\centering
\resizebox*{2in}{2in}{\rotatebox{0}{\includegraphics{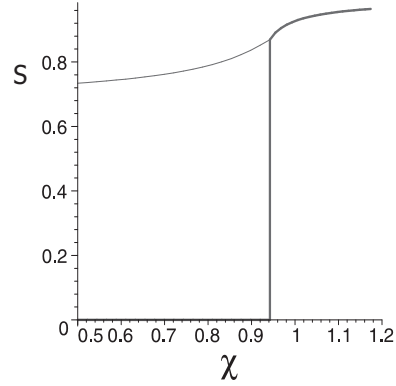}}}
\par}
\caption{Scalar nematic order parameter $S$ as function of the
Flory-Huggins parameter $\chi$. The volume curve is the thick line
and the thin curve is the nematic branch of the isotropic and
nematic coexisting phases of Fig.~\ref{Fig1}. All the parameters
are the same as in Fig.~\ref{Fig1}.} \label{Fig:order_param2}
\end{figure}

\begin{figure}
{\par\centering
\resizebox*{2.8in}{4in}{\rotatebox{0}{\includegraphics{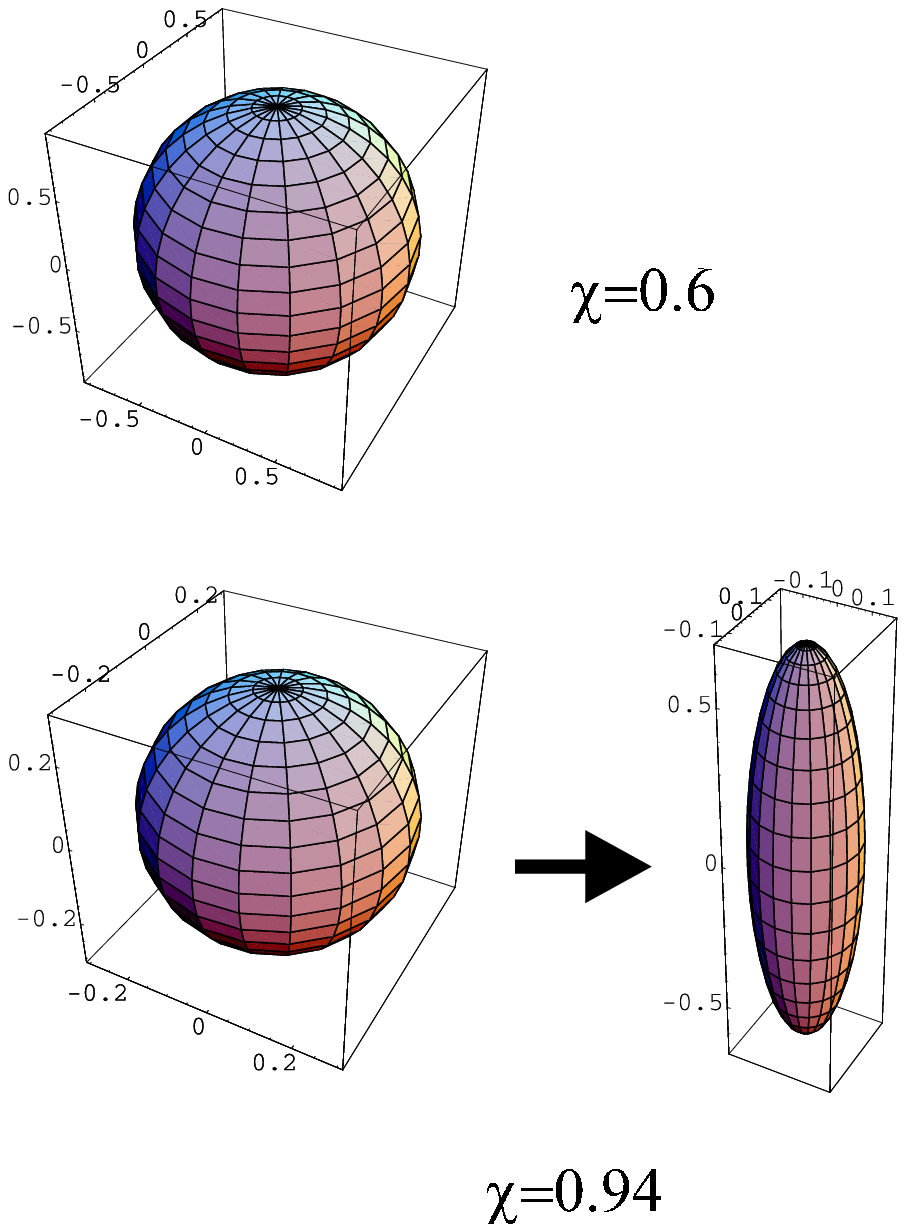}}}
\par}
\caption{Schematic picture of the extension of
the chains inside the gel as characterized by the
 strains $\lambda_{\parallel}$ and $\lambda_{\perp}$,
corresponding to the parameters of Fig.~\ref{Fig1}. The
upper picture corresponds to $\chi=0.6$ in the isotropic phase,
and the two lower pictures correspond to the isotropic and nematic
phases which coexist when the collapse occurs at $\chi^* \simeq
0.94$.  After the collapse, the resultant strains are
$\lambda_{\parallel} \sim 0.72$ and $\lambda_{\perp} \sim 0.17$.}
\label{Fig:schema}
\end{figure}

In Fig.~\ref{Fig1} the phase diagram of the gel is given in two
sets of coordinates: in (a) in terms of $\phi$ vs. $\chi$ and (b)
in terms of the osmotic pressure $\Pi$ vs. $\chi$. Thick solid
lines are isobars of constant osmotic pressure for $\Pi=0.005$,
and $\Pi=0$. The most remarkable feature of these isobars is the
discontinuity in volume fraction (collapse) at a particular value
of $\chi$ defined as $\chi^*$. The isobar $\Pi=0$ is called the
volume curve, and it delimits the region of absolute stability of
the gel in the absence of applied stress. Collapse on the volume
curve occurs at $\chi^* \simeq 0.94$. We emphasize that this
collapse is fundamentally different from the collapse of an
isotropic gel discussed by Tanaka \cite{tanaka1}. In our case, the
collapse is driven by the first-order nematic-isotropic transition
as can be seen in Fig.~\ref{Fig:order_param2}, which shows the
discontinuous change of the nematic order parameter $S$ along the
volume curve. Note that $S \approx 1$ in the nematic phase, which
means that the rods inside the gel are highly ordered. This is
consistent with experimental observations on nematic elastomers \cite{finkelman}.
As shown in Fig.~\ref{Fig:schema}, the
development of nematic order in the gel results in anisotropy in
the strains.

In Fig.~\ref{Fig1}a, thin solid lines delimit the
isotropic-nematic coexistence: the lower line $\phi_A(\chi)$ is
the boundary of the isotropic phase and the upper line
$\phi_B(\chi)$ is the boundary of the nematic phase. As mentioned
in Sec.~\ref{sec:equil_swelling}, if the external condition
changes rapidly and the system is quenched to a particular point
within the coexistence region, macroscopic phase separation will
occur. The volume fractions of the rods in the two phases are
$\beta \phi_A$ and $\beta \phi_B$, respectively. They are both of
the order of $a/b$, which is smaller than the maximum volume
fraction of the rods $\beta$. Note that the coexistence region
widens as $\chi$ increases. This may be attributed to the
interaction between the rods becoming less repulsive as $\chi$
increases. A similar feature has been pointed out by Flory \cite{flory1}
and Flory and Warner \cite{flory2} in a different
system consisting of a solution of rods in the absence of any
elastic network. Note also that a qualitatively similar behavior
has been predicted theoretically \cite{warren,lekkerkerker} and
observed experimentally \cite{fraden} in solutions of rods where
the interaction between rods is attractive and is induced by
depletion interaction.

\subsection{Nematic gels with mobile rods: model 2}
\label{model1}

\begin{figure}
{\par\centering
\resizebox*{2.5in}{2.5in}{\rotatebox{0}{\includegraphics{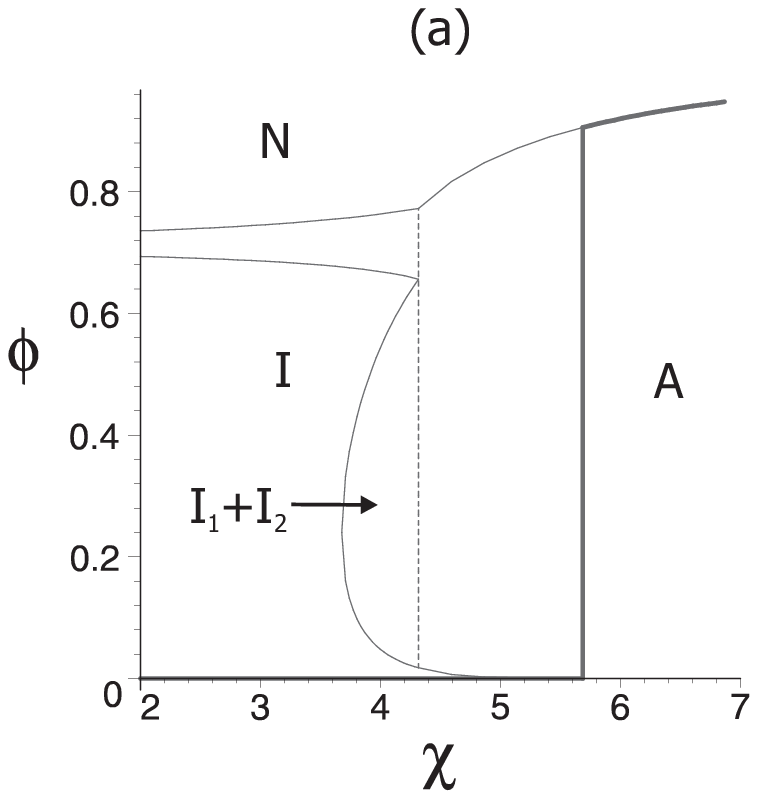}}}
\resizebox*{2.5in}{2.5in}{\rotatebox{0}{\includegraphics{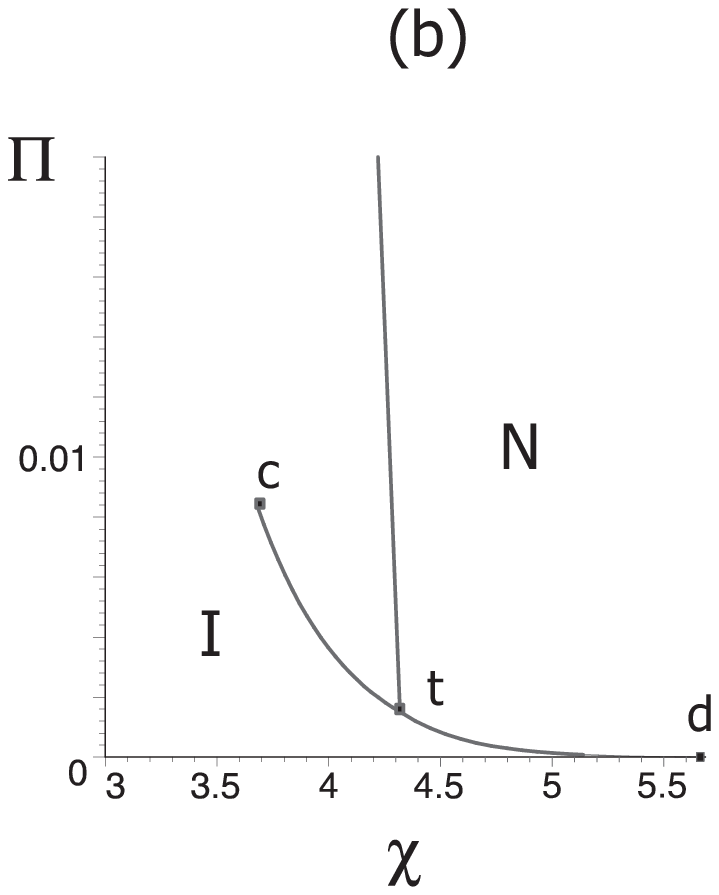}}}
\par}
\caption{Phase diagram for model 2 in the coordinates: (a) volume
fraction of the network $\phi$ vs. Flory-Huggins parameter $\chi$
and (b) osmotic pressure $\Pi$ vs. $\chi$. The thick solid line in
(a) is the volume curve and thin solid lines are coexistence lines
of isotropic-nematic and isotropic-isotropic phases. In the region
denoted $A$ in (a), the osmotic pressure of the gel is negative.
Solid lines in (b) represent first order transitions. The
parameters are $b/a=8$, $\beta=0.628$, $\nu_c =
  5 \cdot 10^{-5}$ and $\phi_0 = 10^{-2}$.
There is a discontinuous volume change (collapse) at $\chi^*
\simeq 5.68$, between an isotropic phase (I) of volume fraction
$\phi_{iso} \simeq 1.1 \cdot 10^{-4}$ and a nematic phase (N) of
volume fraction $\phi_{nem}=0.9$. The coexistence region of the
two isotropic gels is denoted $I_1+I_2$ and terminates at a
critical point denoted $c$ and located at $\chi_c \simeq 3.67$. At
$\chi_t \simeq 4.31$ there is a triple point represented by the
dotted line in (a) and by the point $t$ in (b), where two
isotropic gels of different volume fractions coexist with a
nematic gel.} \label{Fig:model1}
\end{figure}

\begin{figure}
{\par\centering
\resizebox*{2.1in}{2.1in}{\rotatebox{0}{\includegraphics{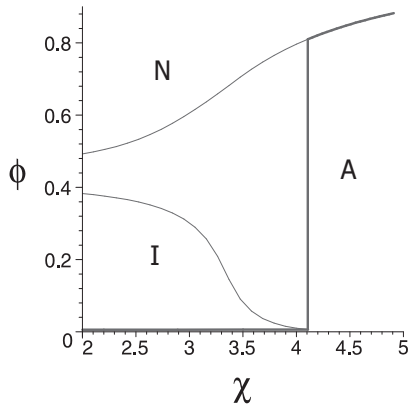}}}
\par}
\caption{For model 2, volume curve (thick line) as a function
 of the Flory-Huggins parameter $\chi$, and coexistence region of
 isotropic-nematic (thin lines). The parameters are
 the same as in Fig.~\ref{Fig:model1} except for $\nu_c$
 which has been increased to  $7 \cdot 10^{-4}$.
 Now the collapse is at $\chi^* \simeq 4.1$.} \label{Fig:nem1}
\end{figure}

\begin{figure}
{\par\centering
\resizebox*{2.5in}{2.5in}{\rotatebox{0}{\includegraphics{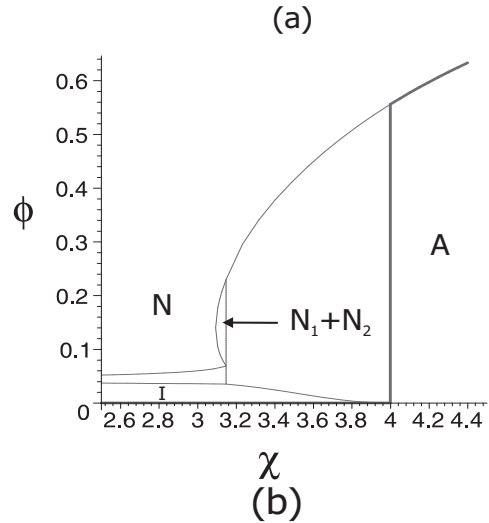}}}
\resizebox*{2.5in}{2.5in}{\rotatebox{0}{\includegraphics{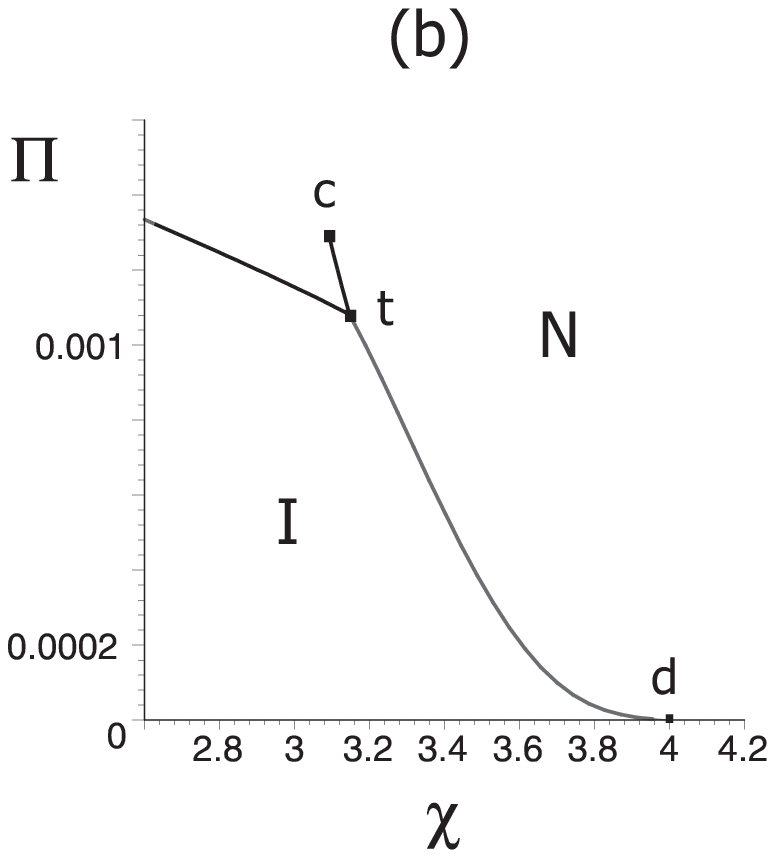}}}
\par}
\caption{Phase diagram for model 2 in the coordinates: (a) $\phi$
vs. $\chi$ and (b) $\Pi$ vs. $\chi$. The parameters are $b/a=100$,
$\beta=0.78$, $\nu_c = 7 \cdot 10^{-6}$ and $\phi_0 = 3 \cdot
10^{-3}$. The collapse occurs at $\chi^* \simeq 4$. The
coexistence region of two nematic gels denoted $N_1+N_2$
terminates at a critical point located at $\chi_c' \simeq 3.1$ and
$\phi_c' \simeq 0.14$. At $\chi_t \simeq 3.15$ there is a triple
point represented as a dotted line in (a) and by the point $t$ in
(b) where two nematic gels with different volume fractions and
nematic order parameters coexist with an isotropic gel.}
\label{Fig:nem2}
\end{figure}

In model 2, the rods are mobile but confined within the gel.
Unlike model 1, the parameters in model 2 are less restricted.
Using a typical value for the shear modulus of an elastomer $\sim
10^{-5}\,\mbox{J/m$^3$}$ \cite{terentjev} and from
Ref.~ \cite{tanaka1}, we estimate that $\nu_c \sim 10^{-5}$ and
$\phi_0 \sim 0.001$. As can be seen in Figs. \ref{Fig:model1} to
\ref{Fig:nem2} with different values of $\nu_c$, $\beta$, and
$b/a$, the phase diagrams of model 2 have two common features
with model 1: the discontinuous collapse and the coexistence of
isotropic-nematic phases.  In addition, model 2 has two
additional interesting features: the coexistence of two isotropic
phases along a line terminating at a critical point and the
coexistence of two nematic phases along another line terminating
at another critical point.

To see the coexistence of two isotropic phases, we expand the
osmotic pressure Eq.~(\ref{osmotic_pressure}) in the isotropic
phase ($\alpha=0$) up to fourth order in $\phi$:
\begin{equation}\label{osmotic_series}
\frac{\Pi v_c}{k_B T}= \delta \phi^2 +
\frac{\phi^3}{3}-\tilde{\nu_c}\phi^{1/3} + \xi \phi,
\end{equation}
where $\delta=\frac{1}{2}-\chi (1-\beta) + 4\beta^2/ \pi$ and
$\tilde{\nu_c}=\nu_c/\phi_0^{1/3}$. The last term $\xi \phi$ in
Eq.~(\ref{osmotic_series}), with $\xi = 4a \beta/\pi b$ comes from
the translational entropy of the rods proportional to $\beta \phi
\ln \phi$ in Eq.~(\ref{F_rod_1}), which is specific to model 2. The
coexistence of two isotropic phases ends at a critical point,
the location of which is found by requiring the first and second
derivative of the osmotic pressure with respect to $\phi$ to be zero
$\Pi'=\Pi''=0$. This critical point located at ($\delta_c$,$\phi_c$) only
exists in fact when $\Pi(\delta_c,\phi_c) \geq 0$. This last condition imposes that
$\xi \geq \xi_e$, with $\xi_e = \frac{4}{9}375^{1/4} \tilde{\nu_c}^{3/4}$.
Therefore as $\xi$ varies, the critical point follows a line which ends
at a critical endpoint when $\xi=\xi_e$ \cite{tanaka1}.
Since $\xi_e > \xi = 0$ in model 1, the volume curve does not have a
critical point in this case and the coexistence between two
isotropic phases is not possible, as found indeed in the previous section.

Equation (\ref{osmotic_series}) was derived by Tanaka for an
isotropic gel with $\xi = \nu_c/2\phi_0$. As discussed in section
\ref{sec:model}, the term proportional to $\xi$ arises from the
additional strain dependent term $\ln \left( \lambda_\parallel^{2}
\lambda_\perp \right)$ in the elastic free energy.  Hence, the
existence of the line of coexistence of two isotropic gels
terminating at a critical point as observed by Tanaka is possible
only if this term is included \cite{tanaka1}. In the absence of
this term, there could only be a {\em continuous} volume change
induced by the change of quality of the solvent. In contrast, we
have found in the case of nematic gels a discontinuous volume
change even without including the controversial strain dependent
term. Thus, this confirms that the physical origin of the collapse
for nematic gels is the isotropic-nematic transition.

Fig.~\ref{Fig:model1} shows the coexistence lines between two
isotropic gels terminating at a critical point at $\chi_c$, in the
case where this critical point is located at $\chi_c  < \chi^*$.
The dotted line in Fig.~\ref{Fig:model1} represents a triple
point, where the two isotropic gels coexist with a nematic gel.
For a larger value of $\nu_c$, we note that either $\chi_c>\chi^*$
or when there is no critical point, and the phase diagram looks
very much like Fig.~\ref{Fig1}, as shown in Fig.~\ref{Fig:nem1}.
Thus, the isotropic-nematic coexistence region has swallowed the
region of coexistence of two isotropic gels. This means that in
this region phase separation into two isotropic phases is
immediately followed by phase separation into a nematic phase and
an isotropic phase.  Note also that in this case, the collapse
occurs at a smaller value of $\chi^* \simeq 4$. This is expected
since the elasticity of the network favors contraction of the gel.
For completeness, the phase diagram of Fig.~\ref{Fig:model1} is
shown in the coordinates osmotic pressure $\Pi$ vs. $\chi$.

When the ratio of the volume fraction of the rods and that of the
network defined as $\beta$ is increased while keeping the other
parameters fixed, the coexistence region of isotropic and nematic
phases moves to lower volume fraction $\phi$. This opens up the
possibility of coexistence of two nematic phases at higher volume
fraction with different volume fractions and different order
parameters. This is indeed the case as shown in
Fig.~\ref{Fig:nem2} for $\beta=0.78$ and $b/a=100$.  The
coexistence region of the two nematic phases ends at a critical
point located at $\chi_c' \simeq 3.1$ and $\phi_c' \simeq 0.14$,
and at a triple point $\chi_t \simeq 3.15$ where two nematic gels
coexist with an isotropic gel. The coexistence of two nematic
phases has been predicted in the literature in systems different
from the one we have considered in this paper. In a solution of
rods in the absence of an elastic network, it was predicted by
Flory \cite{flory1,flory2}, and more recently it was predicted in
Ref.~\cite{warner1} in a system composed of a nematogenic network
immersed in a nematogenic solvent. The coexistence of two nematic
phases is a novel feature of our model, which is more subtle than
in previous studies of nematic gels since in our case only the
network has nematogenic properties. The phase diagram of
Fig.~\ref{Fig:nem2} is shown in the coordinates osmotic pressure
$\Pi$ vs. $\chi$. As compared with Fig.~\ref{Fig:model1}, the
critical point is now on the other side of the line of coexistence
of the isotropic and nematic phases.

\section{Summary and Conclusion}
\label{summary}

In this paper, we have studied the phase behavior and the collapse
of a lyotropic nematic gel using Onsager theory to describe the
isotropic-nematic transition. Upon decreasing the quality of the
solvent sufficiently, we find that a nematic gel always undergoes
a discontinuous volume change, even when the rods embedded in the
gel are not physically linked to the network. The discontinuous
volume change is accompanied by an isotropic-nematic transition,
and for this reason it is of a different nature than the one
discussed by Tanaka for isotropic gels. We have discussed the
possible phase diagrams that these systems should exhibit; in
particular, we have shown that three distinct coexistence phases
are possible, {\em i.e.} isotropic/isotropic (I/I),
isotropic/nematic (I/N), and nematic/nematic (N/N). The I/N
coexistence is always present, the I/I coexistence and N/N
coexistence may be present depending on the precise value of the
parameters.

Our model uses the Onsager free energy for rods in an isotropic
solvent to describe rods in a polymer gel.  We considered two models:
Mmodel 1, in which rods are a part of the network and model 2, in which
the rods are free to move within the free space of the gel but not permitted to
escape from it.  In model 2, we retained the full translational entropy of the
Onsager theory, whereas in model 1, we set it equal to zero since the rods formed
a part of the polymer network.  Thus, in both models, we overestimate the rotational
entropy of the rods and we overestimate rod translational entropy in model 2.
It would be interesting to consider models that provide more realistic
treatments of these entropies.  Our treatment also neglects
completely the random torques and forces that the randomly cross-linked
network exerts on the rods.  These random fields are definitely present, and
they may make it difficult to create an aligned lyotropic nematic gel in the
laboratory.

We hope, nevertheless, that this paper will motivate experimental work on gels
with rods embedded in them. These experiments are important
because of the potential applications of these materials: these
gels are in some sense a new type of sensors, which respond to a
change in osmotic pressure by a change of volume. In particular,
our work is relevant to some experiments which are now undertaken
using crosslinked diblock copolymers as rigid rods embedded in a
polymer network \cite{discher}.\\

\acknowledgement{We acknowledge M. Warner for stimulating discussions at an
early stage of this paper, Ranjan Mukhopadhyay, P. Dalhaimer, and
D. Discher for the many fruitful discussions. This work was
supported in part by the NSF under Grant No. DMR00-9653 (D.L. and
T.C.L.) and by NIH under Grant No. HL67286 (A.W.C.L.).}

\end{document}